\documentstyle{article}

\def\1ad{\mbox{\normalsize $^1$}}
\def\2ad{\mbox{\normalsize $^2$}}
\def\3ad{\mbox{\normalsize $^3$}}
\def\4ad{\mbox{\normalsize $^4$}}
\def\5ad{\mbox{\normalsize $^5$}}
\def\6ad{\mbox{\normalsize $^6$}}
\def\7ad{\mbox{\normalsize $^7$}}
\def\8ad{\mbox{\normalsize $^8$}}
\def\makefront{\vspace*{1cm}\begin{center}
\def\newtitleline{\\ \vskip 5pt}
{\Large\bf\titleline}\\
\vskip 1truecm
{\large\bf\authors}\\
\vskip 5truemm
\addresses
\end{center}
\vskip 1truecm
{\bf Abstract:}
\abstracttext
\vskip 1truecm}
\newcommand{\eref}[1]{(\ref{#1})}

\def\beq{\begin{equation}}
\def\eeq{\end{equation}}
\def\bea{\begin{eqnarray}}
\def\eea{\end{eqnarray}}
\def\ra{\rightarrow}
\def\a{\alpha}
\def\b{\beta}

\def\l{\lambda}
\def\f{\phi}

\def\e{\varepsilon}

\def\G{\Gamma}

\begin {document}

\begin{flushright}
Preprint Padova, DFPD 97/TH\\
December 1997\\
\end{flushright}

\def\titleline{Covariant actions for chiral supersymmetric bosons
\footnote{To appear in the proceedings of the conference "{\it
Quantum Aspects of Gauge Theories, Supersymmetry and Unification}",
Neuchatel University, Neuchatel, Switzerland, 18-23 September 1997.}}

\def\authors{Kurt Lechner}
\def\addresses
{Dipartimento di Fisica, Universit\`a di Padova,\\
Istituto Nazionale di Fisica Nucleare, Sezione di Padova,\\
Via F. Marzolo 8, I--35131 Padova}
\def\abstracttext
{
We review a recently developed covariant Lagrangian formulation 
for $p$--forms with (anti)self--dual field--strengths and present
its extension to the supersymmetric case. As explicit examples
we construct covariant Lagrangians for six--dimensional models
with $N=1$ rigid and curved supersymmetry.
}
\makefront

\section{Introduction}

Chiral bosons are described by $p$--form gauge potentials $B_p$
whose curvatures $H_{p+1}=dB_p$ satisfy, as equation of motion,
a Hodge (anti)self--duality condition in a space--time with 
dimension $D=2(p+1)$. In space--times with 
Minkowskian signature $\eta_{ab}=(1,-1,\cdots,-1)$ the 
self--consistency of such an equation restricts $p$ to even values
and hence the relevant dimensions are $D=2,6,10,\ldots$

Such fields populated superstring and supergravity theories, and more
recently M theory, from their very beginning. Two dimensional chiral 
bosons (scalars) are basic ingredients in string theory, six--dimensional
ones belong to the supergravity-- and tensor--multiplets in $N=1$,
$D=6$ supergravity theories and are necessary to complete the 
$N=2$, $D=6$ supermultiplet of the M-theory five--brane; finally 
a ten--dimensional chiral boson appears in $IIB$, $D=10$
supergravity.

A peculiar feature of the (manifestly Lorentz covariant) self--duality
equation of motion of those fields is that a manifestly Lorentz
invariant lagrangian formulation for them was missing for long time.
The absence of a Lorentz invariant action from which one can derive
the equations of motion leads in principle to rather problematic
situations e.g. the conservation of the energy--momentum tensor
is not guaranteed a priori and the coupling to gravity can not be performed
via the usual minimal coupling.

For previous attempts in facing this problem and for a more detailed 
discussion of the problematic aspects involved, see in particular
\cite{probl}.

Recently a new manifestly Lorentz--invariant lagrangian 
approach for chiral bosons has been proposed in \cite{PST,PST3}. 
The most appealing features of this approach are the introduction 
of {\it only one} single scalar auxiliary field, its natural
compatibility with all relevant symmetries, in particular
with diffeomorphisms and with $\kappa$--invariance \cite{M5},
and its general validity in all dimensions $D=2(p+1)$ with $p$ even.
Another characteristic feature of this approach is the appearance 
of two new local bosonic symmetries: one of them reduces the
scalar auxiliary field to a non propagating "pure gauge" field and
the other one  reduces the second order equation of motion
for the $p$--form to the first order (anti)self--duality
equation of motion.

A variant of this approach allowed to write manifestly duality
invariant actions for Maxwell fields in four dimensions
\cite{PSTDUAL} and to construct a covariant effective action 
for the M theory five--brane \cite{M5}. On the other
hand, the actions obtained
through the non manifestly covariant approach developed 
in \cite{schw} can be regarded as gauge fixed versions of the
actions in \cite{M5,PSTDUAL} where the scalar auxiliary field has been
eliminated.

The coupling of all these models with chiral bosons to gravity can be
easily achieved since the new approach is manifestly covariant
under Lorentz transformations; as a consequence it is
obvious that the two above mentioned  bosonic symmetries, which are
a crucial ingredient of the new approach, are compatible with
diffeomorphism invariance. To test the general
validity of the approach, it remains to establish its compatibility
with global and local supersymmetry. This is the aim of the
present talk.

In the next section we review the covariant method, for definiteness,
for chiral two--forms in six dimensions. In section three we 
test its compatibility with supersymmetry by writing a covariant
action for the most simple cases, i.e. the rigid tensor supermultiplet
and the free supergravity multiplet in six dimensions. Section
four is devoted to some concluding remarks and to a brief discussion
of the general case i.e. the supergravity multiplet coupled to an
arbitrary number of tensor multiplets and super Yang--Mills multiplets.

The general strategy developed in this paper extends in a rather
straightforward way to two and ten dimensions. Particularly interesting 
is the case of $IIB$, $D = 10$ supergravity whose covariant 
action we hope to present elsewhere. The bosonic part of this
action has already been presented in \cite{IIB}.

For more details on the results presented here and for more
detailed references, see \cite{DLT}.

\section{Chiral bosons in six dimensions: the general method}

In this section we present the method for a chiral boson
in interaction with an external or dynamical gravitational
field in six dimensions. To this order we introduce sechsbein one--forms
$e^a = d x^m {e_m}^a(x)$. With  $m,n =0,\ldots,5$ we indicate 
curved indices and with $a,b=0,\ldots,5$ we indicate tangent
space indices, which are raised and lowered with the flat
metric $\eta_{ab}=(1,-1,\cdots,-1)$.

To consider a slightly more general self-duality condition for
interacting chiral bosons we introduce the two-form potential 
$B$ and its generalized curvature three--form $H$ as
$$
H=dB+C\equiv {1\over 3!}e^a e^b e^c H_{cba},
\label{forms}
$$
where $C$ is a three-form which depends on the fields to which 
$B$ is coupled, such as the graviton, the gravitino and so on, 
but not on $B$ itself. The free (anti)self--dual boson 
will be recovered for $C=0$ and $e_m{}^a=\delta_m{}^a$.

The Hodge--dual of the three--form $H$ is again a three--form
$H^*$ with components $H^*_{abc} = \frac{1}{3!} \e_{abcdef} H^{def}.$  
The self--dual and anti self--dual parts of $H$ are defined respectively
as the three--forms $H^{\pm} \equiv \frac{1}{2} (H \pm H^*)$.
The equations of motion for interacting chiral bosons in supersymmetric 
and supergravity theories, as we will see in the examples worked out in
the next section, are in general of the form $H^{\pm}=0,$
for a suitable three--form $C$ whose explicit expression is
determined by the model.

To write a covariant action which eventually gives rise to
the equation $H^{\pm}=0$
we introduce as new ingredient the scalar auxiliary field $a(x)$
and the one--form
\beq
v={1\over \sqrt{-\partial_c a \partial^c a }}\, da\equiv e^b v_b.
\eeq
In particular we have $v_b={\partial_b a\over 
\sqrt{-\partial_c a \partial^c a}}$ 
and $v_bv^b=-1$. Using the vector $v^b$, to the 
three--forms $H,H^*$ and $H^\pm$ we can then associate two-forms $h,h^*$ 
and $h^\pm$ according to 
$$
h_{ab}=v^cH_{abc}, \qquad h={1\over 2} e^a e^b h_{ba},
$$
and similarly for $h^*$ and $h^\pm$.

The action we search for can now be written equivalently in one of the
following two ways
\beq
\label{S0}
S_0^\pm = \pm\int \left(v h^{\pm} H + {1\over 2} dB C\right)
                   = \int d^6x\sqrt{g}\left({1\over 24}H_{abc}H^{abc} 
         +{1\over 2}h_{ab}^\pm h^{\pm ab}\right) \pm\int {1\over 2}dBC.
\eeq 
$S_0^+$ will describe anti self--dual bosons ($H^+=0$)
and $S_0^-$ self--dual bosons ($H^-=0$). 
The last term, $\int dBC$, is of the Wess--Zumino type and is absent 
for free chiral bosons. 

What selects this form of the action are  essentially 
the local symmetries it possesses. Under a general variation of the fields
$B$ and $a$ it varies, in fact, as
\beq
\label{dS0}
\delta S_0^\pm = \pm2\int \left(vh^\pm d\delta B + 
{v\over \sqrt{-\partial_c a \partial^c a }} h^{\pm}h^{\pm} d\delta a\right).
\eeq
From this formula it is rather easy to see that $\delta S^\pm_0$ vanishes 
for the following three bosonic transformations,  with transformation
parameters $\Lambda$ and $\psi$, which are one--forms, and $\varphi$ 
which is a scalar:
\bea\label{bos}
&I)&\qquad \delta B=d\Lambda,\qquad \delta a =0\nonumber\\
&II)&\qquad \delta B= -{2h^\pm\over \sqrt{-\partial_c a \partial^c a }}\,
\varphi,\qquad \delta a =\varphi\nonumber\\
&III)&\qquad \delta B=\psi da ,\qquad \delta a =0.
\eea

For what concerns $I)$ and $III)$ invariance of the action is actually 
achieved also for finite transformations. This fact will be of some importance
below. 

The transformation $I)$ represents just the ordinary gauge 
invariance for abelian two--form gauge potentials.
The symmetry $II)$ implies that $a(x)$ is an 
auxiliary field which does, therefore,  not correspond to a propagating 
degree o freedom\footnote{Notice however that, since the action becomes 
singular in the limit of a vanishing or constant $a(x)$, the gauge 
$d a(x) = 0$ is not allowed.}. Finally, the symmetry $III)$ eliminates
half of the propagating degrees of freedom carried by $B$ and allows to
reduce the second order equation of motion for this field to the desired
first order equation, i.e. $H^{\pm}=0$. To see this we note that the equations
of motion for $B$ and $a$, which  can be read from \eref{dS0}, 
are given respectively by
\beq
d\left(vh^\pm\right)=0\label{emb},\qquad
d\left({v\over \sqrt{-\partial_c a \partial^c a }}h^\pm h^\pm\right)=0.
\eeq 
First of all it is straightforward to check that the $a$--equation is 
implied by the $B$-equation, as expected, while the general solution of 
the $B$--equation is given by $vh^\pm={1\over 2}d\tilde{\psi}da$,
for some one--form $\tilde{\psi}$. On the other hand, under a (finite) 
transformation $III)$ we have $\delta\left(vh^\pm\right)={1\over 2}d\psi 
da$ and therefore, choosing $\psi=-\tilde\psi$, 
we can use this symmetry to reduce the $B$-equation 
to $vh^\pm=0$. But $vh^\pm=0$ amounts to $h^\pm=0$, and this equation, 
in turn, can easily be seen to be equivalent to $H^\pm=0$, the 
desired chirality condition.

This concludes the proof that the actions $S_0^\pm$ describe indeed
correctly the propagation of chiral bosons.

In a theory in which the $B$ field is coupled to other dynamical 
fields, for example in supergravity theories, we can now conclude
that the complete action has to be of the form 
$$
S=S_0^\pm+S_6,
$$
where $S_6$ contains the kinetic and interaction terms for the fields
to which $B$ is coupled. To maintain the symmetries $I)$--$III)$
one has to assume that those fields are invariant under these
transformations and, moreover, that $S_6$ is independent of $B$ and $a$.

For more general chirality conditions describing self--interacting 
chiral bosons, like e.g. those of the Born--Infeld type, see ref. \cite{PST3}. 

To conclude this section we introduce two three--form fields,
$K^\pm$, which will play a central role in the next section
due to their remarkable properties. They are defined as
\beq\label{k}
K^\pm\equiv H+2vh^\mp
\eeq
and are uniquely determined by the following peculiar properties:
i) they are (anti) self--dual: $K^{\pm*} = \pm K^{\pm}$;
ii) they reduce to $H^\pm$ respectively  if  $H^\mp= 0$;
iii) they are invariant under the symmetries $I)$ and $III)$, 
         and under $II)$ modulo the field equations \eref{emb}.
These fields constitute therefore a kind of off--shell generalizations
of $H^\pm$.

\section{$N=1$, $D=6$ supersymmetric chiral bosons}

In this section we illustrate the compatibility of the general 
method for chiral bosons with supersymmetry in the six--dimensional case
by means of two examples:  the first concerns the free
tensor supermultiplet in flat space--time and the second concerns 
pure supergravity. The strategy
developed in these examples admits natural extensions to more general
cases \cite{M5,IIB,DLT}.

\vskip0.5truecm\noindent
{\it 1) Free tensor multiplet.}
\vskip0.5truecm

An $N=1,D=6$  tensor multiplet is made out of an antisymmetric tensor 
$B_{[ab]}$, a symplectic Majorana--Weyl 
spinor $\l_{\a i}$ ($\a = 1,\ldots,4; i = 1,2$) and a real scalar $\f$.
The equations of motion for this multiplet and its on--shell susy transformation
rules are well known. The scalar obeys the free Klein--Gordon equation,
the spinor the free Dirac equation and the $B$--field the self--duality
condition $H^-=0,$ where $H=dB$, which means that in this case we have $C=0$.

The on-shell supersymmetry transformations, with rigid transformation
parameter $\xi^{\a i}$, are given by 
\bea
\label{susy}
\delta_\xi \f &=& \xi^i \l_i, \nonumber\\
\delta_\xi \l_{ i} &=& \left( \G^a \partial_a \f +
\frac{1}{12} \G^{abc}H_{abc}^+ \right)\xi_i,\nonumber\\
\delta_\xi B_{ab} &=& - \xi^i \G_{ab} \l_i.
\eea
The $USp(1)$  indices 
$i,j$ are raised and lowered according to
$ K_i = \e_{ij} K^j,  K^i = - \e^{ij} K_j, $ where $\e_{12} =1$ and
the $\Gamma^a$ are $4\times 4$ Weyl matrices.

Since the equations of motion are free our ansatz for the action, which 
depends now also on the auxiliary field $a$, is 
\beq\label{SH}
S=S_0^-+S_6
=- \int v h^- H +{1\over 2}\int d^6x \left(\l^i \G^a \partial_a \l_i +
\partial_a \f \partial^a \f \right).
\eeq
This action is invariant under the symmetries $I)$--$III)$ if we assume
that $\f$ and $\l$ are invariant under these transformations.

For what concerns supersymmetry we choose first of all for $a$ the 
transformation $\delta_\xi a=0,$
which is motivated by the fact that $a$ is non propagating and
does therefore not need a supersymmetric partner. Next we should 
find the off--shell generalizations of \eref{susy}. For dimensional
reasons only $\delta_\xi\l$ allows for such an extension. To find
it we compute the susy variation of $S_0^-$, which depends only on $B$ and
$a$, as
$$
\delta_\xi S_0^-=-2\int vh^-d\delta_\xi B=-\int K^+d\delta_\xi B
$$
in which the self-dual field $K^+$, defined in the previous section,
appears automatically. Since $\delta_\xi S_0^-$  should be cancelled by
$\delta S_6$ this suggests to define the off--shell
susy transformation of $\l$ by making the simple replacement
$H^+\ra K^+$, i.e.
$$
\delta_\xi \l_{ i}\ra \bar\delta_\xi \l_{ i} = \left( \G^a \partial_a \f +
\frac{1}{12} \G^{abc}K_{abc}^+ \right)\xi_i.
$$
With this modification it is now a simple exercise to show that
the action \eref{SH} is indeed invariant under supersymmetry.
The relative coefficients of the terms in the action are actually
fixed by supersymmetry.

The {\it general rules} for writing covariant actions for supersymmetric
theories with chiral bosons, 
which emerge from this simple example, are the following.
First one has to determine the on--shell susy transformations of the 
fields and their equations of motion, in particular one has to determine
the form of the three-form $C$. For more complicated theories 
this can usually be done most conveniently using superspace techniques.
The off--shell extensions of the susy
transformation laws are obtained by substituting in the transformations
of the fermions $H^\pm\ra K^\pm$. The action has then to be written
as $S_0^\pm+S_6$ where the relative coefficients of the various terms 
in $S_6$ have to be determined by susy invariance. The field $a$, finally,
is required  to be supersymmetry invariant. 

\vskip0.5truecm\noindent
{\it 2) Pure supergravity.}
\vskip0.5truecm

The supergravity multiplet in six dimensions contains the graviton, 
a gravitino and an 
antisymmetric tensor with anti--selfdual (generalized) field strength. 
The graviton is described by the vector--like vielbein $e^a = dx^m 
{e_m}^a$, the gravitino by the spinor--like one--form $e^{\a i} = 
dx^m {e_m}^{\a i}$ and the tensor by the two--form $B$. 

The supersymmetry transformations of these fields and their equations 
of motion, obtained from the superspace approach \cite{DFR}, are conveniently
expressed in terms of a super--covariant differential,  
$D=d+\omega$, with respect to a super--covariant Lorentz connection one--form 
$\omega^{ab} = dx^m \omega_{m}{}^{ab}$. This connection is defined by
$d e^a + e^{b}{\omega_b}^a = - e^i\G^a e_i,$
and results in the metric connection augmented by the standard gravitino
bilinears. 

Among the equations of motion we recall the generalized 
anti--selfduality condition for $B$. This reads 
$H^+=0,$ where now
$$
H=dB+ \left(e^i\G_a  e_i\right) e^a,
$$
meaning that in this case the three--form $C$ is non vanishing
being given by $C=\left(e^i\G_a  e_i\right) e^a.$

The on--shell supersymmetry transformations of $e^a$, $e^{\a i}$ and $B$ 
\cite{DFR}, with local transformation parameter $\xi^{\a i} (x)$, are given by 
\bea
\label{susyi}
\delta_\xi e^a &=& -2 \xi^i \G^a e_i, \qquad
\delta_\xi e^{\a i} = D\xi^{\a i} - \frac{1}{8} \xi^{\b i}  e^a 
(\G^{bc})_\b{}^\a H^-_{abc}, \label{28b}\\
\delta_\xi B &=& -2 (\xi^i \G_a e_i) e^a, \qquad
\delta_\xi a = 0.
\eea

According to our general rule, in the gravitino transformation we
have to make the off--shell replacement $H^{-}\rightarrow K^{-}$ obtaining
$$
\delta_\xi e^{\a i}\rightarrow
\bar\delta_\xi e^{\a i}= D\xi^{\a i} - \frac{1}{8} \xi^{\b i}  e^a 
(\G^{bc})_\b{}^\a K^-_{abc}.
$$

In the above relations we added the trivial transformation law for the 
auxiliary field $a$. As it stands, this trivial
transformation law does not seem to preserve the susy algebra in that
the commutator of two supersymmetries does not amount to a translation.
On the other hand it is known that the supersymmetry algebra closes on 
the other  symmetries of a theory; in the present case it is easily seen 
that the anticommutator of two susy transformations on the $a$ 
field closes on the  bosonic transformations $II)$. 

The covariant action for pure $N = 1$, $D = 6$ supergravity can 
now be written as $S=S_0^+ +\int L_6$, where
\bea\label{azsu} 
     S_0^+&=& \int \left(v h^+ H + {1\over 2} dB C\right)\\
     L_6 &=&
 \frac{1}{48} \e_{a_1 \ldots a_6 } e^{a_1} e^{a_2} e^{a_3} e^{a_4} 
R^{a_5a_6}  - \frac{1}{3} e^{a_1}e^{a_2}e^{a_3} (De^i\G_{a_1 a_2 a_3}e_i). 
\eea
For convenience
we wrote the term $S_6$ as an integral of a six--form, $L_6$. This
six--form contains just the Einstein term, relative to the 
super--covariantized spin connection 
$R^{ab}=d\omega^{ab} +\omega^a{}_c\omega^{cb}$, and the kinetic term for
the gravitino. The relative coefficients are fixed by susy invariance.
In this case $S_0^+$ contains also the couplings of
$B$ to  the gravitino and the graviton.

This action is invariant under the symmetries $I)$--$III)$ because
$L_6$ is independent of the fields $B,a$ and we assume the graviton 
and the gravitino to be invariant under those transformations.

The evaluation of the supersymmetry variation of $S$ is now a merely
technical point and can indeed be seen to vanish. In particular, as
in the previous example, the susy variation of $S^+_0$ depends on the 
fields $B,a$ only through the combination $K^-$ and these contributions
are cancelled by the gravitino variation, justifying again our rule for 
the modified susy transformation rules for the fermions.

\section{Concluding remarks}

The covariant lagrangians presented in this talk for six--dimensional
supersymmetric chiral bosons admit several extensions. 
The lagrangian for $n$ tensor multiplets coupled to the supergravity 
multiplet, which involves $n+1$ mixed (anti)self--duality conditions,
has been worked out in \cite{DLT}. The introduction of hyper multiplets,
on the other hand, does not lead to any new difficulty. The coupling to 
Yang--Mills fields in the presence of $n$ tensor multiplets requires some
caution. In this case it turns out that an action, and therefore a 
consistent classical theory, can be constructed only if the $n+1$ 
two--forms can be arranged such that only one of them carries a
Chern--Simons correction while the $n$ remaining ones have as invariant
field strength $dB^{(n)}$.

In conclusion the covariant method illustrated in this talk appears
compatible, at the classical level, with all relevant symmetries explored
so far, in particular with supersymmetry. 

Among the open problems at the quantum level one regards the existence
of a covariant quantization procedure. A quantum consistency check of
the covariant method for chiral bosons coupled to gravity, which has still 
to be performed, consists in a perturbative computation of the Lorentz 
anomaly and in a comparison with the result predicted by the index theorem.

\paragraph{Acknowledgements.}
It is a pleasure for me to thank G. Dall'Agata and M. Tonin for 
their collaboration on the results presented in this talk. 
I am also grateful to I. Bandos, P. Pasti and D. Sorokin for their interest
in this subject and for useful discussions. This work was supported by the 
European Commission TMR programme ERBFMPX-CT96-0045.

\end{document}